\documentclass[aps,prd,preprint,superscriptaddress,showpacs,nofootinbib]{revtex4-1}
\usepackage{graphicx}
\usepackage{subfigure}
\usepackage{multirow}
\usepackage{ulem}
\usepackage{color}
\definecolor{My_red}        {cmyk}{0.00,1.00,1.00,0.20}

\usepackage{slashed}
\usepackage{dcolumn}

\newcommand{\bmat}{\left(\begin{array}}
\newcommand{\emat}{\end{array}\right)}
\newcommand{\beq}{\begin{equation}}
\newcommand{\eeq}{\end{equation}}

\def\bwt{\begin{widetext}}
\def\ewt{\end{widetext}}
\def\be{\begin{equation}}
\def\ee{\end{equation}}
\def\bea{\begin{eqnarray}}
\def\eea{\end{eqnarray}}
\def\bean{\begin{eqnarray*}}
\def\eean{\end{eqnarray*}}
\def\bary{\begin{array}}
\def\eary{\end{array}}
\def\bit{\begin{itemize}}
\def\eit{\end{itemize}}

\def\su5u1{SU(5) \times U(1)}
\def\fsu5u1{SU(5) \times U(1)'}
\def\so10{SO(10)}
\def\sq20{SO(10) \times SO(10)}

\def\nn{\nonumber}

\def\bwt{\begin{widetext}}
\def\ewt{\end{widetext}}
\def\be{\begin{equation}}
\def\ee{\end{equation}}
\def\bea{\begin{eqnarray}}
\def\eea{\end{eqnarray}}
\def\bean{\begin{eqnarray*}}
\def\eean{\end{eqnarray*}}
\def\bary{\begin{array}}
\def\eary{\end{array}}
\def\bit{\begin{itemize}}
\def\eit{\end{itemize}}

\def\su5u1{SU(5) \times U(1)}
\def\fsu5u1{SU(5) \times U(1)'}
\def\so10{SO(10)}
\def\sq20{SO(10) \times SO(10)}

\usepackage[centertags]{amsmath}
\usepackage{amssymb}

\begin{document}

\title{An $\boldsymbol{SU(6)}$ GUT Origin of the TeV-Scale Vector-like Particles Associated with the 750 GeV Diphoton Resonance}

\author{Bhaskar Dutta}

\affiliation{
Mitchell Institute for Fundamental Physics and Astronomy,
Department of Physics and Astronomy, Texas A\&M University, College Station, TX 77843-4242, USA }

\author{Yu Gao}

\affiliation{
Mitchell Institute for Fundamental Physics and Astronomy,
Department of Physics and Astronomy, Texas A\&M University, College Station, TX 77843-4242, USA }

\author{Tathagata Ghosh}

\affiliation{
Mitchell Institute for Fundamental Physics and Astronomy,
Department of Physics and Astronomy, Texas A\&M University, College Station, TX 77843-4242, USA }

\author{Ilia Gogoladze}

\affiliation{
Bartol Research Institute, Department of Physics and Astronomy,
University of Delaware, Newark, DE 19716, USA }

\author{Tianjun Li}

\affiliation{Key Laboratory of Theoretical Physics and
Kavli Institute for Theoretical Physics China (KITPC),
Institute of Theoretical Physics, Chinese Academy of Sciences,
Beijing 100190, P. R. China}

\affiliation{
  School of Physical Sciences, University of Chinese Academy of Sciences,
  Beijing 100049, P. R. China
}

\affiliation{
School of Physical Electronics, University of Electronic Science and Technology of China,
Chengdu 610054, P. R. China
}

\author{Joel W. Walker}

\affiliation{Department of Physics, Sam Houston State University, Huntsville, TX 77341, USA}

\date{\today}

\begin{abstract}
We consider the $SU(6)$ GUT model as an explanation for the diphoton final state excess,
where the masses of all associated particles are linked with a new symmetry breaking scale.
In this model, the diphoton final states arise due to loops involving three pairs of new vector-like particles
having the same quantum numbers as down-type quarks and lepton doublets. These new vector-like fermions are embedded
alongside the SM fermions into minimal anomaly-free representations of the $SU(6)$ gauge symmetry. The $SU(6)$ symmetry is broken
to the Standard Model times $U(1)_X$ at the GUT scale, and masses for the vector-like fermions arise at the TeV scale only after the
residual $U(1)_X$ symmetry is broken. The vector-like fermions do not acquire masses via breaking of the SM symmetry at the EW scale.
The field which is responsible for the newly observed resonance belongs to the $\bar{6}_H$ representation. 
The dark matter arises from the SM singlet fermion residing in $\bar{6}$ and is of Majorana type.
We explicitly demonstrate gauge coupling unification in this model, and also discuss the origin of neutrino masses.
In addition to the diphoton final states, we make distinctive predictions for
other final states which are likewise accessible to the ongoing LHC experimental effort.
\end{abstract}


\preprint{MI-TH-1614}

\maketitle

\section{Introduction}

Recently, both the ATLAS and CMS collaborations have reported an excess of events in the diphoton channel
at a reconstructed invariant mass of about 750 GeV. This excess is visible in both the 13 TeV only~\cite{ATLAS_diphoton:2015, CMS_diphoton:2015}
and $13+8$ TeV~\cite{ATLAS_diphoton:Moriond,CMS_diphoton:Moriond} LHC data analyses.
The ATLAS collaboration reports a local signal significance of $3.9 \, \sigma$ from an integrated luminosity of 3.2 ${\rm fb}^{-1}$ at 13 ~TeV,
and about 1.9 $\sigma$ from 20.3 ${\rm fb}^{-1}$ of 8 TeV data. The CMS collaboration likewise reports a local signal significance
of 3.4 $\sigma$ from combined luminosities of 3.3 and 19.7 ${\rm fb}^{-1}$ at 13 and 8 TeV, respectively.
CMS finds the observed significance to be maximized for a narrow decay width $\Gamma_S/M_S\lesssim 10^{-4}$, while the ATLAS data
is reported to favor a larger width with $\Gamma_S/M_S\sim 0.06$.  The collection of more data is necessary to clarify the status of the observed excess and the associated decay width.

A straightforward approach to explaining the diphoton excess is the introduction of a Standard Model
(SM) singlet $S$ with a mass of 750~GeV accompanied by multiplets of vector-like particles~\cite{diphoton:Strumia,diphoton:Zurek,diphoton:first}.
With vector-like particles in the loops, the singlet $S$ can be produced via gluon fusion,
and can likewise decay into a diphoton pair. The vector-like particles can solve the vacuum stability
problem~\cite{Gogoladze:2010in}.

In contrast to masses of the chiral fermions of the SM,
the masses of the vector-like quark and lepton
are not tied to the electroweak (EW) scale,
since they do not arise from the breakdown of the SM gauge symmetry.
The natural question that arises is whether we may introduce a new symmetry that can be broken down at the TeV scale to generate masses
for the vector-like fermions. In this paper, we employ the gauge group $SU(6)$, placing the SM quarks, leptons and also
the new vector-like quarks and leptons into anomaly free $15$, $\bar{6}$ and $\bar{6}$ representations.
$SU(6)$ is a subgroup of the anomaly-free exceptional group $E_6$, i.e., $E_6\supset SU(6)\times SU(2)$,
implying that our results may likewise be embedded within the context of an $E_6$ model.
In our scenario, the $SU(6)$ gauge symmetry is broken down to $SU(3)_C\times SU(2)_L \times U(1)_Y \times U(1)_X $
at the GUT scale, and the residual $U(1)_X$ gauge symmetry is broken at the TeV scale by the VEV of SM singlets belonging to $\bar{6}_H$ and $21$, producing
vector-like masses for 3 generations of new down-type quarks and lepton doublets. The ratio of vector-like quark and lepton masses $M_D/M_L$ gets fixed.
The 750 GeV resonance arises from a scalar field 
which is the SM singlet within a $\bar{6}_H$ of $SU(6)$.
The SM is subsequently broken at the weak scale.
In addition to the new vector-like quarks and leptons, the
adoption of fundamental representations of $SU(6)$ also naturally implies two sets of SM singlets carrying the $X$ charge
for each generation which develop Majorana masses $\sim$ TeV. One set of particles interacts with the SM leptons and
will be responsible for the lightness of the neutrino masses via a seesaw mechanism~\cite{seesaw}.
The lightest component of the second set will be a dark matter (DM) candidate.
We also investigate the gauge coupling unification in this model.

In Section 2 we detail the $SU(6)$ model and discuss the vector-like particle masses, neutrinos, and DM.
In Section 3 we discuss the gauge unification and GUT symmetry breaking.
In Section 4 we discuss the diphoton excess and predict cross-sections for various other final states in the context of the $SU(6)$ model.
In Section 5 we conclude.

\section{The $SU(6)$ Model}

In order to explain the observed excess of events around 750 GeV in the diphoton channel, we need to introduce
new particles beyond the SM spectrum at a low scale.  Particles with fermionic degrees of freedom are slightly better motivated than scalars,
since their loop-induced contributions are larger, in general.  Also, mass stability is much easier to explain in the fermionic case.
It is very natural to ask about an underlying mechanism for the introduction of new vector-like particles into the spectrum, and
whether the necessary fields are an arbitrary choice or one governed by the enlarged symmetry structure of some grand unified theory (GUT).
In particular, we are interested in the question of what dynamics may protect the TeV scale masses of these particles from
GUT or Planck scale corrections.

Whereas the SM fermions neatly fit into representations of $SU(5)$ or $SO(10)$, the minimal group structure which
provides natural unification of the SM chiral fermions with additional particles transforming as vector-like particles
under the SM gauge symmetry is the $SU(6)$ GUT.
The smallest anomaly-free set of chiral representations which fulfill this purpose (for one particle generation) in the $SU(6)$ GUT are
 \begin{eqnarray}
2\times \bar{6} + 15.
\label{eq1}
\end{eqnarray}
The $\overline{6}$ and $15$ dimensional representations decompose under the the gauge symmetry as follows.
 \begin{eqnarray}
15 &=& (q, u^c, e^c) \oplus (\bar{L}_{15}(1, 2, 1/2),~ \bar{D}_{15}(3,1,-1/3)) \\
\bar 6 &=& (d^{c}, l) \oplus N^{\prime} \\
\bar 6^{\prime} &=& (L_{6}(1,2,-1/2), D_{6}(\overline 3, 1,1/3)) \oplus N
\label{eq2}
\end{eqnarray}
Here we are using the common notation $q$, $u^c$, $e^c$, $d^c$, and $l$ for the SM fermions.
$D_6$, $L_6$, $\bar D_{15}$, and $\bar L_{15}$ are vector-like particles arising from ${\bar 6}_i$
and $15_i$, $N $ and $N'_i$ are singlet fermions.  For simplicity, we presently consider
$d^c$, and $l$ to be elements of the $\bar 6$ representation,
while the new vector-like particles are in $\bar 6^{\prime}$.
However, as we shall elaborate later, the physical SM $d^c$, and $l$ will actually
arise from a superposition of $\bar 6$ and $\bar 6^{\prime}$.

It is interesting to note that the additional vector-like particles can only obtain mass once the rank of the $SU(6)$
gauge symmetry is broken.  Specifically, the vector-like particles are chiral under the residual $U(1)_X$ subgroup
of $SU(6)$, and will remain massless until this symmetry is broken.
We consider the scenario where $SU(6)$ is broken at the GUT scale to $SU(3)_c\times SU(2)_L \times U(1)_Y \times U(1)_X$,
and the $U(1)_X$ breaking scale is around a TeV.  Thus, the $SU(6)$ GUT may facilitate a well-defined vector-like particle
spectrum with a common mass scale around a TeV.

The Yukawa sector in our model has the following form.
 \begin{eqnarray}
{\cal L} &=& 15\cdot 15 \cdot 15_H + (\bar 6 + \bar 6^{\prime})\cdot 15 \cdot \bar 6_H +
{\bar 6}\cdot {\bar 6} \cdot 15_H +  (\bar 6 + \bar 6^{\prime})\cdot 15 \cdot  {\bar 6^{\prime}}_{H} + {\rm h.c.}
\label{eq3}
\end{eqnarray}
We have suppressed family and gauge indices for simplicity.
In order to generate the SM fermion masses and mixing, as well as the vector-like particle masses,
we need to introduce the following $SU(6)$ representation for the Higgs field.
 \begin{eqnarray}
15_H + \bar 6_H + \bar 6^{\prime}_H
\label{eq4}
\end{eqnarray}
The up-type Higgs lives in $15_H$, the down-type Higgs lives in $\bar 6_H$,
and the SM singlet field which can break the extra $U(1)$ symmetry is in $\bar 6^{\prime}_H$.
We need to introduce two different representations, i.e. $15_H + \bar 6_H$, for the SM Higgs fields in order
to have realistic quark and lepton masses.  The reason for this is that taking only a single
representation $\bar 6_H$ (or $15_H$) will lead to a non-renormalizable coupling for one of the Yukawas~\cite{Shafi:2001iu}.
For instance, the non-renormalizable coupling in $15\cdot 15 \cdot \bar 6_H \cdot \bar 6^{\prime}_H/M$
would lead to a very small effective Yukawa coupling to the up type quarks, since the singlet field responsible for $U(1)$
symmetry breaking takes a vacuum expectation value (VEV) at the TeV scale in our scenario.

The first term in Eq.~(\ref{eq3}) characterizes the up-type quark mass matrix and mixing.
The second term does the same for down-type quark and charged leptons.
The assumption that $d^c$, and $l$ live only in the $\bar 6$ representation
implies $b$-$\tau$ Yukawa coupling unification at the GUT scale.
This condition is problematic, since the $b$ and $\tau$ Yukawa couplings meet each other around $10^{5}$ GeV in the SM
if we run under the renormalization group equations (RGEs) from low scale to high.
However, considering instead that $d^c$, and $l$ are superpositions of $\bar 6$ and $\bar 6^{\prime}$ breaks the
$b$-$\tau$ Yukawa unification condition such that there is no conflict with experimental data.
This mixing between the SM particles and the vector-particle particles is likewise helpful~\cite{delAguila:2008pw}
to explain the BNL muon $g-2$ data~\cite{BNL}.
The third term in Eq.~(\ref{eq3}) provides for Dirac neutrino masses.
The final term in Eq.~(\ref{eq3}) generates vector-like masses for the new particles when the SM singlet component in $6^{\prime}_H$
generates a VEV which breaks the $U(1)_X$ symmetry around the TeV scale,
and moreover provides the scalar particle candidate ($S$) responsible for the observed 750~GeV resonance.\\

\vspace{0.5 in}

\noindent{\it{Neutrino Mass and Dark Matter}}
\\

In order to avoid cosmological constraints on the number of degrees of freedom for massless particles~\cite{Neff},
we need to generate a large mass for each of the $N$ and $N^{\prime}$ fields, one of which we will consider
as a right-handed neutrino which interacts with the SM like $l$.
The singlets $N$ and $N^{\prime}$ can acquire Majorana mass from the interaction
\begin{eqnarray}
({\bar 6}\cdot \bar{6} + {\bar 6}\cdot \bar{6^{\prime}} +{\bar 6^{\prime}}\cdot \bar{6^{\prime}})\cdot 21.
\label{eq5}
\end{eqnarray}
The $21$ dimensional representation of $SU(6)$ contains single $S'$ under the SM gauge symmetry \cite{Slansky:1981yr}.
The VEV for this field is also associated with the TeV scale and can be responsible for breaking the $U(1)_X$ symmetry.
$N^{\prime}$ also interacts with $l$ and generates
a Dirac mass, cf. Eq. (\ref{eq3}). We thus have all the necessary
ingredients for realization of a type-I seesaw mechanism for neutrino masses and mixing.

The lightest of the three generations of singlet Majorana type fields $N$ can be the DM candidate,
if also lighter than $L$'s and $D$'s.



\section{$\boldsymbol{SU(6)}$ GUT symmetry breaking and gauge coupling unification}

As described previously, we are considering the symmetry breaking $SU(6) \rightarrow SU(3)_c\times SU(2)_L \times U(1)_Y \times U(1)$
at the GUT scale. In order to realize this process we require at least two scalar adjoint representations ($\Phi_1+ \Phi_2$) in the theory.
For simplicity, we assume that $\Phi_1$ and $\Phi_2$ have a global $Z_2$ symmetry.
In this case, the most general renormalizable potential involving only $\Phi_1$ and $\Phi_2$ has the following form.
\begin{eqnarray}
  V&=&-\frac{M^2_1}{2}{\rm Tr}[\Phi_1^2] + \frac{\lambda_1}{4}{\rm Tr}[\Phi_1^2]^2 +\frac{\lambda_2}{4}{\rm Tr}[\Phi_1^4] -\frac{M^2_2}{2}{\rm Tr}[\Phi_2^2] +
\frac{\lambda_3}{4}{\rm Tr}[\Phi_2^2]^2 +\frac{\lambda_4}{4}{\rm Tr}[\Phi_2^4] \nonumber \\
  &+&\frac{\lambda_5}{2}{\rm Tr}[\Phi_1^2]{\rm Tr}[\Phi_2^2] +\frac{\lambda_6}{2}{\rm Tr}[\Phi_1^2\Phi_2^2] + \lambda_7{\rm Tr}[\Phi_1^2\Phi_2^2]
  \label{eq6}
\end{eqnarray}
For simplicity we further assume that $\lambda_6\ll\lambda_i$ and $\lambda_7\ll\lambda_i$ ($i\neq$ 6 or 7).
One possible VEV configuration of the $\Phi_1$ and $\Phi_2$ fields is
\begin{eqnarray}
  \Phi_1 &=& \frac{1}{2{\sqrt {3}}}  {\rm diag} \left(1,~1,~1,~-1,~-1,~-1\right) V_{\Phi_1}~,~\, \\
  \Phi_2 &=& \frac{1}{2{\sqrt {6}}} {\rm diag} \left(1,~1,~1,~-2,~-2,~1\right) V_{\Phi_2}~,~\,
  \label{eq7}
\end{eqnarray}
where
\begin{eqnarray}
 V_{\Phi_1}= \frac{6M_2^2\lambda_5 - M^2_1(6\lambda_3+\lambda_4)}{72\lambda_5^2-3(4\lambda_1+\lambda_2)(6\lambda_3+\lambda_4)}~,\\
 V_{\Phi_2}=\frac{M_2^2(4\lambda_1 + \lambda_2)-4M_1^2\lambda_5}{(4\lambda_1+\lambda_2)(6\lambda_3+\lambda_4)-24\lambda_5^2}~.
 \label{eq8}
\end{eqnarray}

\begin{figure}[t]
\centering \includegraphics[angle=0, width=12cm]{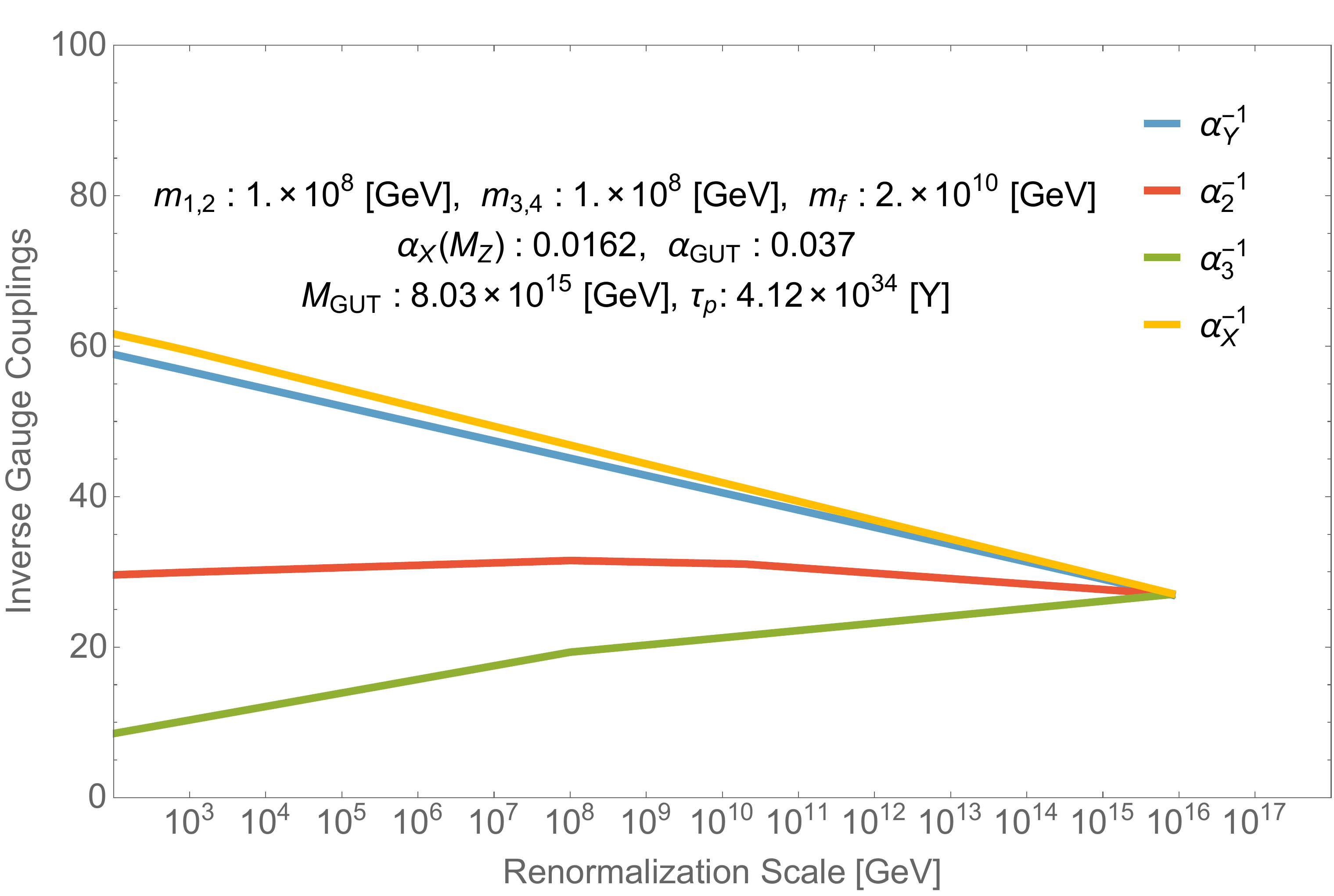}
\vspace{-0.3cm} \caption{ Gauge coupling evolution with
mass scales $m_{1,2} = m_{3,4} = 1\times10^8$~GeV
for the adjoint triplet and octet scalars (2 each)
and $m_f^2 = 2\times10^{10}$~GeV for a single
fermionic weak triplet.  Three pairs of vector-like $(5,\bar{5})$
are additionally introduced, with a split mass hierarchy of
$1$~TeV and $600$~GeV for the colored and
non-colored field components, as well as a second Higgs doublet
and a pair of light single scalars at 750~GeV.
This corresponds to scenario $D$ of Table~\ref{tab:unificationbms}.}
\label{fig:unification_D}
\end{figure}

After GUT symmetry breaking, various components of the $\Phi_1$ and $\Phi_2$ scalar
multiplets obtain different masses, as in Table~\ref{tab:scalarmasses}.
At the low scale we also have vector-like particles transforming under the SM gauge symmetry.
\begin{eqnarray}
3\times \left[ (1, 2, 1/2)+ (1,2,-1/2)+ (3,1,-1/3) + (\overline 3, 1,1/3) \right]
 \label{eq9}
\end{eqnarray}
These multiplets are from the fermionic ($15+\bar 6$) representations of the $SU(6)$ GUT,
constituting full $5+\bar 5$ dimensional representations of the $SU(5)$ subgroup.
As shown in Eq.~(\ref{eq5}), these vector-like particles will obtain a common mass once the $\bar 6^{\prime}$
field develops a VEV for its sixth element, breaking the additional $U(1)_X$ gauge symmetry.
Since all vector-like particles from Eq.~(\ref{eq8}) have the same mass, they
will not change the relative slopes of RGE running for the gauge couplings at one-loop level and will induce only a slight modification at two-loop level.
So, as shown in Figure~\ref{fig:unification_D}, gauge coupling unification is obtained by a suitable choice of the
$\lambda_1$ coupling, as reflected in the physical masses of the particles in Table~\ref{tab:scalarmasses}.
In order to have a light Higgs doublet at the low scale, a fine-tuning procedure is required,
as is characteristic of any non-supersymmetric GUT.

\begin{table}[]
\begin{center}
\def\arraystretch{1.5}%
\scalebox{0.80}{
\begin{tabular}{|c|c|c|c|c|c|c|c|c|c|}
\hline
 $m_1^2$& $m_2^2$ & $m_3^2$& $m_4^2$ &  $m_5^2$ &  $m_6^2$ & $m_7^2$ & $m_8^2$  & $m_9^2$ &  $m_{10}^2$ \\
 $(8,1,0)_1$& $(8,1,0)_2$ & $(1,3,0)_1$ & $(1,3,0)_2$& $(3,1, 1/3)$ & $(\bar 3,1,-1/3)$ & $(1,2,1/2)$ & $(1,2,-1/2)$ &$(3,2, 5/6)$& $(\bar 3, 2, -5/6)$ \\
 \hline
  $2\lambda_4 V^2_2$& $4\frac{\lambda_5^2}{\lambda_4} V^2_1$ & $2\lambda_4 V^2_2$ & $2\lambda_2 V^2_1$ & $(\lambda_6+\lambda_7)V^2_1$ & $(\lambda_6+\lambda_7)V^2_1$ & $4(\lambda_6+\lambda_7)V^2_2$ & $4(\lambda_6+\lambda_7)V^2_2$ & $M_{GUT}$ & $M_{GUT}$\\
 \hline
 \end{tabular}}
\caption{The masses of physical scalar particles from the $\Phi_1$ and $\Phi_2$ adjoint multiplets after $SU(6)$ gauge symmetry breaking.}
\label{tab:scalarmasses}
\end{center}
\end{table}

We have studied evolution of the $SU(3)_c\times SU(2)_L \times U(1)_Y \times U(1)$ gauge couplings under the renormalization group at the second loop,
including leading feedback between the single loop evolution of the top, bottom and charm Yukawa couplings and the SM gauge sector.  The
relevant RGEs are
\begin{equation}
\frac{d \alpha_i}{dt} = \frac{b_i \alpha_i^2}{2\pi}
 +\frac{\alpha_i^2}{8\pi^2}
\left[~ \sum_{j=1}^4 B_{ij}  \,\alpha_j \right]~,
\label{eq:mssmrge}
\end{equation}
where $\alpha_i \equiv (\alpha_Y,\alpha_2,\alpha_3,\alpha_X)$, suppressing printing of the Yukawa sector.
The associated one-loop ($b_i$) and two-loop ($B_{ij}$) $\beta$-function coefficients are given in Eqs.~(\ref{eq:b},\ref{eq:B}),
where indices $I,II$ respectively denote the field content active around the
the TeV scale, and above the intermediate ($\sim 10^{10}$ GeV) scale.
\begin{equation}
b^{I}=\left(\frac{31}{5},-1,-5,\frac{403}{60}\right)
\quad,\quad
b^{II}=\left(\frac{31}{5},\frac{5}{3},-3,\frac{403}{60}\right)
\label{eq:b}
\end{equation}
\begin{equation}
\begingroup
\renewcommand*{\arraystretch}{1.5}
B^{I}=\begin{pmatrix}
\frac{243}{50}& \frac{63}{10}&12&\frac{77}{50}
\cr
\frac{21}{10} &\frac{65}{2}& 12&\frac{47}{30}
\cr
\frac{3}{2}&\frac{9}{2}&12&\frac{3}{2}
\cr
\frac{77}{50}&\frac{47}{10}&12&\frac{1631}{100}
\end{pmatrix}
\quad,\quad
B^{II}=\begin{pmatrix}
\frac{243}{50}& \frac{63}{10}&12&\frac{77}{50}
\cr
\frac{21}{10} &\frac{547}{6}& 12&\frac{47}{30}
\cr
\frac{3}{2}&\frac{9}{2}&96&\frac{3}{2}
\cr
\frac{77}{50}&\frac{47}{10}&12&\frac{1631}{100}
\end{pmatrix}
\endgroup
\label{eq:B}
\end{equation}
Specifically, the matter content in region $I$ consists of three generations the fermionic fields described Eq.~(\ref{eq1}), i.e
the SM plus $3\times(5, \overline{5})$, as well as the light Higgs plus additional scalars corresponding
to a second electroweak Higgs doublet and two SM singlets from the $\overline{6}$ and $21$ representations.
Masses for colored and non-colored components of the new vector-like particles are fixed at 600~GeV and 1~TeV, respectively.
This scale is favored by the diphoton analysis and this splitting is consistent with that suggested
by running of the relevant Yukawa couplings under the renormalization group.
Since the candidate for the diphoton resonance is included as one of the new scalars, we will choose to assign the new light scalars a common mass of 750~GeV.
In region $II$, we activate two pair each of scalars in the $SU(2)_L$ adjoint scalar triplet and $SU(3)$ adjoint scalar octet, as well as
a single fermionic weak triplet.  The generation of these scalar masses is described in Table~\ref{tab:scalarmasses}, and we will carry
over the notation $m_{1,2}$ and $m_{3,4}$ for the octet and triplet, respectively.  We will simplify to a common mass for each set,
which may be interpreted as a geometric mean.  The mass of the fermionic triplet will be denoted as $m_f$.

Table \ref{tab:unificationbms} reports the induced low-energy value of the $U(1)_X$ coupling $\alpha_X(M_Z)$,
the grand unified coupling $\alpha_{\rm GUT}$, and mass scale $M_{\rm GUT}$, as well as the corresponding dimension-six
proton lifetime $\tau_p$ for four examples of the renormalization group flow.
The unification solution is not greatly affected by small variation of the vector-like mass scale within the physical window,
nor even by omission of these fields (except for a reduction in $\alpha_{\rm GUT}$).
Essentially similar results are obtained with the further mutual inclusion of
one pair each of adjoint scalars carrying the quantum numbers of the right-handed down-quark conjugate and the left-handed lepton doublet at the intermediate scale.
Near GUT scale threshold corrections from scalar fragments, e.g. with quantum numbers of the quark doublet, likewise do little to alter the essential features described.
The first selected scenario $A$ omits the fermionic adjoint triplet, while including one pair each of the octet and triplet scalar adjoints.
It is found that the GUT unification scale is unacceptably light unless the triplet mass is quite low.
Pushing $m_{3,4}$ all the way down to one TeV sets an upper bound of $M_{\rm GUT} \le 5\times10^{15}$~GeV
for this field content.  Triple unification is then achieved for $m_{1,2} \simeq 3\times 10^8$~GeV.
Scenario $B$ introduces additionally a single fermionic weak triplet,
and imposes the constraint of degenerate mass scales $m_{1,2} = m_{3,4} = m_f^2$.
It is found that strict unification near $3\times10^{15}$ GeV
is induced if the new fields are placed at an intermediate scale, around $3\times10^9$~GeV.
The GUT scale in each of the prior scenarios remains somewhat light,
suggesting overly-rapid decay of the proton, with a dimension-six lifetime on the order of $10^{33}$~years.  Given that the GUT scale varies inversely
with the scalar adjoint mass, scenario $C$ is designed to investigate the maximal offset which may achieved relative to the field content of scenario $B$.
If the scalar masses are pushed down to around 10~TeV, then the unification scale moves up to around $10^{17}$~GeV, extending proton decay beyond the reach
of foreseeable experimental searches.  However, it is not phenomenologically necessary to consider such an extreme splitting.  Mild splitting between a minimal
configuration of fields at the intermediate mass scale (with or without inclusion of near-GUT threshold corrections) is sufficient to acceptably elevate the
unification scale.  Scenario $D$ is selected to demonstrate such a physically optimal possibility, taking $m_{1,2} = m_{3,4} = 1\times10^8$~GeV
and $m_f^2 = 2\times10^{10}$~GeV, lifting the unification scale to around $8\times10^{15}$~GeV, and extending
the proton lifetime to a safe yet testable range around $4\times10^{34}$~years.
The scenario $D$ unification is depicted in Figure~\ref{fig:unification_D}.
A stable prediction is made for the low-energy value of the $SU(6)$-normalized coupling $\alpha_{X}$.  Given that the
leading one-loop beta-coefficient $b_4 = 403/60$ is very similar to that of the SM hypercharge
(this is a rather generic feature of $U(1)'$ subgroups from $E_6$ embeddings reflecting
the fact that the number of particles which do not form GUT multiplets is small), the slope of their
running is almost degenerate, and a value $\alpha_X(M_Z) \simeq 0.016$ is to be expected at the
$Z$-boson mass, or a value $\alpha_X({\rm TeV}) \simeq 0.017$ at the TeV scale.

\begin{table}[t]
\begin{center}
\def\arraystretch{1.5}%
\setlength\tabcolsep{6pt}
\scalebox{1.0}{
\begin{tabular}{|c|c|c|c|c|c|c|c|}
\hline
scenario & $m_{1,2}$ & $m_{3,4}$ & $m_f^2$ & $\alpha_{X}(M_Z)$ & $\alpha_{\rm GUT}$ & $M_{\rm GUT}$ & $\tau_p~{\rm [Y]}$ \\
\hline
A & $3\times10^8$ & $1\times10^3$ & NA & 0.016 & 0.037 & $5\times10^{15}$ & $7\times10^{33}$ \\ \hline
B & $3\times10^9$ & $3\times10^9$ & $3\times10^9$ & 0.016 & 0.036 & $3\times10^{15}$ & $1\times10^{33}$ \\ \hline
C & $1\times10^4$ & $1\times10^4$ & $1\times10^{11}$ & 0.016 & 0.041 & $1\times10^{17}$ & $2\times10^{39}$ \\ \hline
D & $1\times10^8$ & $1\times10^8$ & $2\times10^{10}$ & 0.016 & 0.037 & $8\times10^{15}$ & $4\times10^{34}$ \\ \hline
 \end{tabular}}
\caption{Masses (in GeV) of the adjoint triplet and octet scalars (2 each) and a single fermionic weak triplet
for four benchmark unification scenarios.  The resulting low-energy $U(1)_X$ coupling,
grand unification scale (in GeV), and coupling, as well as the
corresponding dimension-six proton lifetime (in years), are also tabulated.}
\label{tab:unificationbms}
\end{center}
\end{table}

We note that there are many other ways to likewise achieve gauge coupling unification in
non-supersymmetric theories~\cite{Amaldi:1991zx}. For instance, one can use the split multiplet mechanism~\cite{Chkareuli:1994ng},
which can explain why we have incomplete multiplets near the GUT scale and facilitate gauge coupling unification around $10^{16}$~GeV.
We emphasize again that the $SU(6)$ gauge group can be embedded into $E_6$, deferring the details of
a reinterpretation of our result in this framework to the appendix.

\section{\label{dPh} The Diphoton Excess}

As described previously, the singlet $S$ of ${\bar 6^{\prime}}_{H}$ is presently considered to
provide the scalar particle responsible for the observed $750$~GeV resonance.
$S$ is coupled to $L$ and $D$ via $\bar 6^{\prime}\cdot 15 \cdot  {\bar 6^{\prime}}_{H}$,
as shown in Eq.~(\ref{eq3}). We thereby get photon, $Z$, $W$, and jet final states.

The leading order decay rate of the resonance $S$ into various diboson final states are given by,
\begingroup\makeatletter\def\f@size{10}\check@mathfonts
\bea
\Gamma(S \rightarrow \gamma \gamma) & = & \frac{M^{3}_S}{64 \pi} \bigg(\frac{e^2}{4 \pi^2}\bigg)^2 \, \Bigg\vert   \sum_{f=D,L} {N_f N_{c}^f Q_{f}^2 \lambda_f} \, \bigg\{ \frac{1}{M_f}\, A_{\frac{1}{2}}(\tau_f) \,  +   \sum_{i=1}^2  \frac{ A_{f}}{2 M_{\tilde{f}}^2} \, A_0(\tau_{\tilde{f}}) \bigg\} \Bigg\vert^2 \, , \nn \\
\Gamma(S \rightarrow ZZ) & = & \frac{M^{3}_S}{64 \pi} \bigg(\frac{1}{4 \pi^2}\frac{g^{2}_2}{c^{2}_W}\bigg)^2 \, \Bigg\vert   \sum_{f=D,L} {N_f N_{c}^f (T_{3}^f - Q_f s^{2}_W)^2 \lambda_f} \, \bigg\{ \frac{1}{M_f}\, A_{\frac{1}{2}}(\tau_f) \,  +   \sum_{i=1}^2  \frac{ A_{f}}{2 M_{\tilde{f}}^2} \, A_0(\tau_{\tilde{f}}) \bigg\} \Bigg\vert^2 \nn\\
 & & \times \, \bigg(1 - 4\frac{ M_{Z}^2}{M_{S}^2} + 6\frac{ M_{Z}^4}{M_{S}^4}  \bigg)  \, \sqrt{1- 4 \frac{ M_{Z}^2}{M_{S}^2} } \, , \nn \\
 \Gamma(S \rightarrow Z \gamma) & = & \frac{M^{3}_S}{32 \pi} \bigg(\frac{1}{4 \pi^2}\frac{e g_2}{c_W}\bigg)^2 \, \Bigg\vert   
 \sum_{f=D,L} {N_f N_{c}^f Q_f (T_{3}^f - Q_f s^{2}_W) \lambda_f} \, \bigg\{ \frac{1}{M_f}\, A_{\frac{1}{2}}(\tau_f) \,  +   \sum_{i=1}^2  \frac{ A_{f}}{2 M_{\tilde{f}}^2} \, A_0(\tau_{\tilde{f}}) \bigg\} \Bigg\vert^2 \nn\\
 & & \times \, \bigg(1 - \frac{ M_{Z}^2}{M_{S}^2}  \bigg)^3 \, , \nn \\
 \Gamma(S \rightarrow W^+W^-) & = & \frac{M^{3}_S}{32 \pi} \bigg(\frac{1}{4 \pi^2}\frac{g^{2}_2}{2}\bigg)^2 \, \Bigg\vert   \sum_{f=D,L} {N_f N_{c}^f \lambda_f} \, \bigg\{ \frac{1}{M_f}\, A_{\frac{1}{2}}(\tau_f) \,  +   \sum_{i=1}^2  \frac{ A_{f}}{2 M_{\tilde{f}}^2} \, A_0(\tau_{\tilde{f}}) \bigg\} \Bigg\vert^2 \nn\\
 & & \times \, \bigg(1 - 4\frac{ M_{W}^2}{M_{S}^2} + 6\frac{ M_{W}^4}{M_{S}^4}  \bigg)  \, \sqrt{1- 4 \frac{ M_{W}^2}{M_{S}^2} }\, , \nn \\
 \Gamma(S \rightarrow g g) & = & \frac{M^{3}_S}{8 \pi} \bigg(\frac{g_{3}^2}{4 \pi^2}\bigg)^2 \, \Bigg\vert   \sum_{f=D} {N_f T_r \lambda_f} \, \bigg\{ \frac{1}{M_f}\, A_{\frac{1}{2}}(\tau_f) \,  +   \sum_{i=1}^2  \frac{ A_{f}}{2 M_{\tilde{f}}^2} \, A_0(\tau_{\tilde{f}}) \bigg\} \Bigg\vert^2 \, , 
 \label{decay_widths}
 \eea
\endgroup
where $N_f = 3$ is the number of copies of ($5, \, \overline{5}$), $N_{c}^f$, being the color-factor, attains a value 3 (1) for $D \, (L)$, $\lambda_f$ are the Yukawa couplings of $f$ with $S$, $A_f$ are the trilinear couplings of $S$ with the SUSY partners of the vector-like fermions, $f=D,L$, if we supersymmetrize our model. $Q_f$ and $T^{f}_3$ are electric charge and third component of the isospin of fermions (and their super-partners whenever they are included in the calculation) respectively. $\sin{\theta_W} \, (\cos{\theta_W})$, where $\theta_W$ is the Weinberg angle, are denoted by $s_W \, (c_W)$ in the above equations. The dynkin index for color triplet $D$, $T_r = 1/2$, is used in the $\Gamma(S \rightarrow g g)$ calculation. Finally the loop functions for spin-1/2 and spin-0 particles are given by,
\bea
A_{\frac{1}{2}}(\tau_f) & = & 2 \int_{0}^1 dx \int_{0}^{1-x} dz \, \frac{1-4xz}{1-xz\tau_f}, \nn \\
A_{0}(\tau_{\tilde{f}}) & = &  \int_{0}^1 dx \int_{0}^{1-x} dz \, \frac{4xz}{1-xz \tau_{\tilde{f}}},
\eea
with $\tau_i = \frac{M_{S}^2}{M_{i}^2}$. Please note that in the decay width calculations involving massive gauge bosons, the effect of gauge boson mass on loop functions have been neglected since they change the loop functions only by $\sim 5 \%$. In addition we also assumed that the mixing between the sparticles ($\tilde{f}_i, \, i=1,2$) is negligible  in the formulas of Eq.~(\ref{decay_widths}). 

A pair of iso-singlet $D$-type quarks can be strongly produced at the LHC and studied in the $H/Z \, b\, +$
anything or $Wt\,+$ anything channels. The current strongest ATLAS bound on $D$-type vector-like quark masses
of $\lesssim 800$ GeV arises from dilepton final state when the $D$ predominantly decays to $Wt$. However the bound relaxes to $\lesssim 650$ GeV if the dominant decay mode is $Hb$ (see~\cite{ATLAS-VLQ} and references therein). The particular BRs for a BP depends on the mixing of $D$ with SM down-type quarks and we can tune the mixing parameters to satisfy the bounds. In contrast, the vector-like leptons $L$ are less likely to be produced at the LHC since
they do not necessarily have large mixings with SM leptons. Hence, they can easily evade the excited lepton searches by CMS~\cite{CMS:VLL}.

We have performed a separate evolution of the $\lambda$ couplings of vector-like fermions in the $3 \times (5, \overline{5})$ representations
between the GUT scale and the scale of the observed resonance $M_S \sim 750$~GeV,
finding attraction toward a fixed point in the vicinity of $\lambda_{L}=0.4$ and $\lambda_{D}=0.7$
that is essentially similar to the result obtained in our previous analysis~\cite{Dutta:SU5} in the context of a pure $SU(5)$ supersymmetric GUT.
We therefore select benchmark masses for the $L$ and $D$ which are broadly consistent with this prediction,
noting that the specific values do not have a significant impact on the gauge unification.
$M_D$ and $M_L$ masses arise due to the VEV of the SM singlet component $\bar{6}'_H$.
The $Z'$ mass associated with the $U(1)_X$, however, arises from the largest VEV of
the SM singlet components of $\bar{6}'_H$, $\bar{6}_H$ and $21$ which is around a TeV.

The diphoton production cross-section at the LHC, assuming the narrow-width approximation, can be written as
\be
  \sigma_{\gamma\gamma}  = \dfrac{K \, \pi^2}{8 M_S}\dfrac{\Gamma(S\rightarrow gg)\Gamma(S \rightarrow \gamma\gamma)}{ \Gamma_S} \times 
 \dfrac{1}{s}\int \limits dx_1 dx_2  f_{g}(x_1) f_g(x_2) \delta \left(x_1 x_2 - \frac{M^{2}_S}{s} \right),
 \label{eq:diphoton_xsec}
\ee
where $\sqrt{s}=13$ TeV, $K$ is the QCD K-factor, $x$ denotes the fraction of each beam's energy carried away by the corresponding gluon,
and $f_g$ is the gluon parton distribution function inside a proton. The total decay width of $S$ is
denoted by $\Gamma_S = \Gamma_{\gamma \gamma}+\Gamma_{Z \gamma}+\Gamma_{Z Z}+\Gamma_{WW}+\Gamma_{gg}$.
We have used the PDFs of {\tt MSTW2008LO}~\cite{MSTW2008} for the
gluon luminosity calculation with the factorization scale set at $M_S$. We evaluated $\alpha_s$ to be 0.092 at our scale of interest but we found that $\alpha$ does not change significantly from its value (0.0078) at $M_Z$. A K-factor of 2.5 is used in our calculation, which is the K-factor for 750~GeV SM-like Higgs~\cite{K-factor}. We also included $\alpha_{s}^4$ correction to $\Gamma_{gg}$, which increases it by a factor of $\sim 1.7$~\cite{Chetyrkin:1997iv}.

We note that the CMS and ATLAS collaboration results disagree to some extent on the experimentally observed width of the resonance.
While ATLAS obtains the highest significance for a large width of $\Gamma_S/M_S = 0.06$, CMS data is fitted better by narrow width
of $\Gamma_S/M_S = 1.4 \times 10^{-4}$. However, the data collected so far is insufficient to support either case convincingly.
The loop induced diphoton and dijet widths are inadequate to account for the $\mathcal{O}(10)$ GeV width required by ATLAS.
Ref.~\cite{Nilles} has recently performed a likelihood analysis to fit 8 and 13 TeV datasets of both CMS and ATLAS experiments.
Interestingly, inclusion of the 8 TeV data lowers the best-fit cross-section by a factor $\sim 2$ and shifts the resonance to $\sim 745$ GeV.
These authors further noticed that a narrow width explanation of the excess reduces the combined significance from $3.9 \, \sigma$ to $3.3 \, \sigma$.
Finally, they conclude that a narrow width resonance between $\sim 730-755$ GeV can be fit by $\sigma_{\gamma \gamma} \sim 1 - 5$~fb
at the $2 \, \sigma$ level (with the best-fit being at 2.6 fb).



\begin{figure}[!t]
\centering \includegraphics[angle=0, width=8cm]{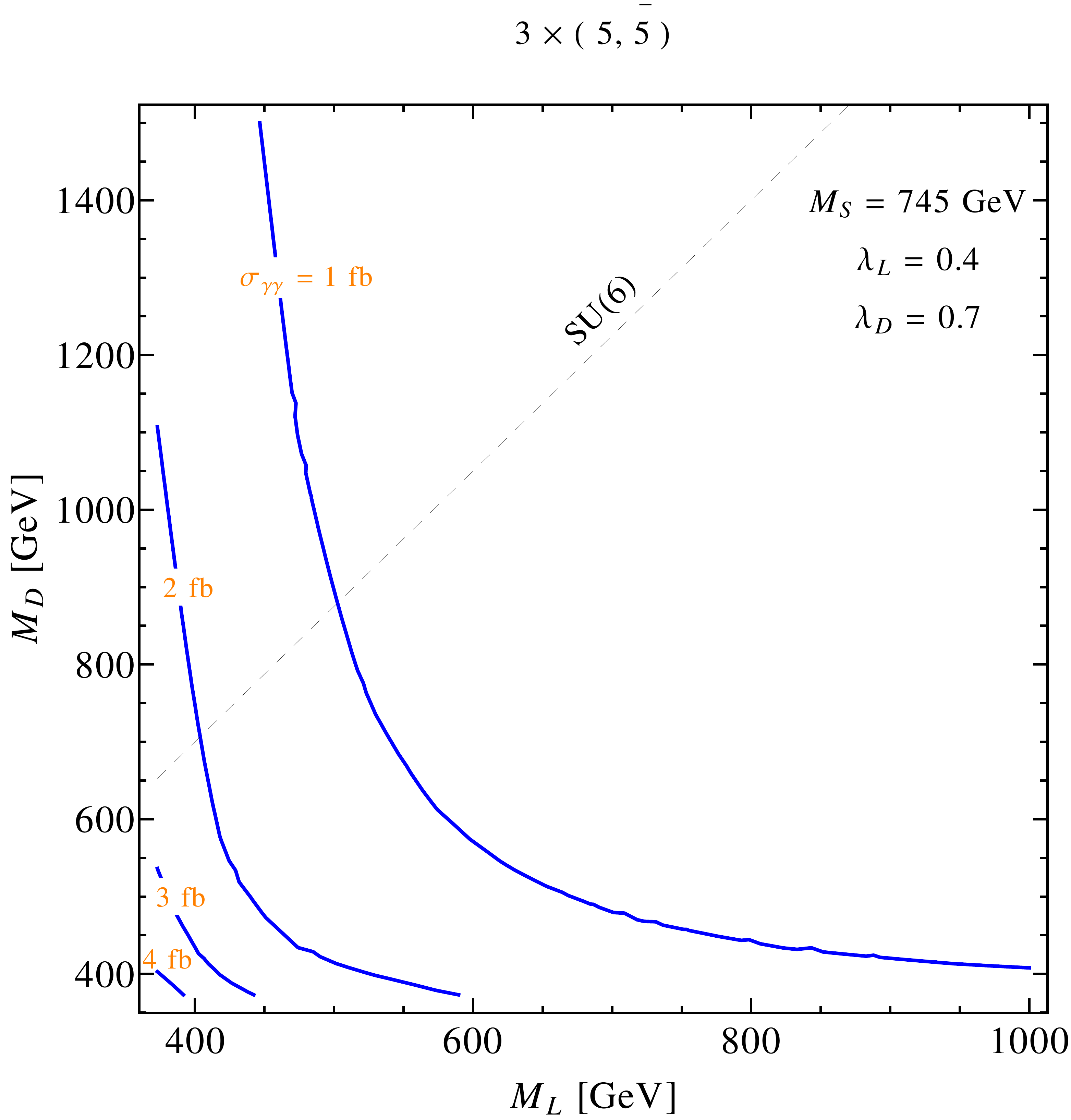}
\centering \includegraphics[angle=0, width=8cm]{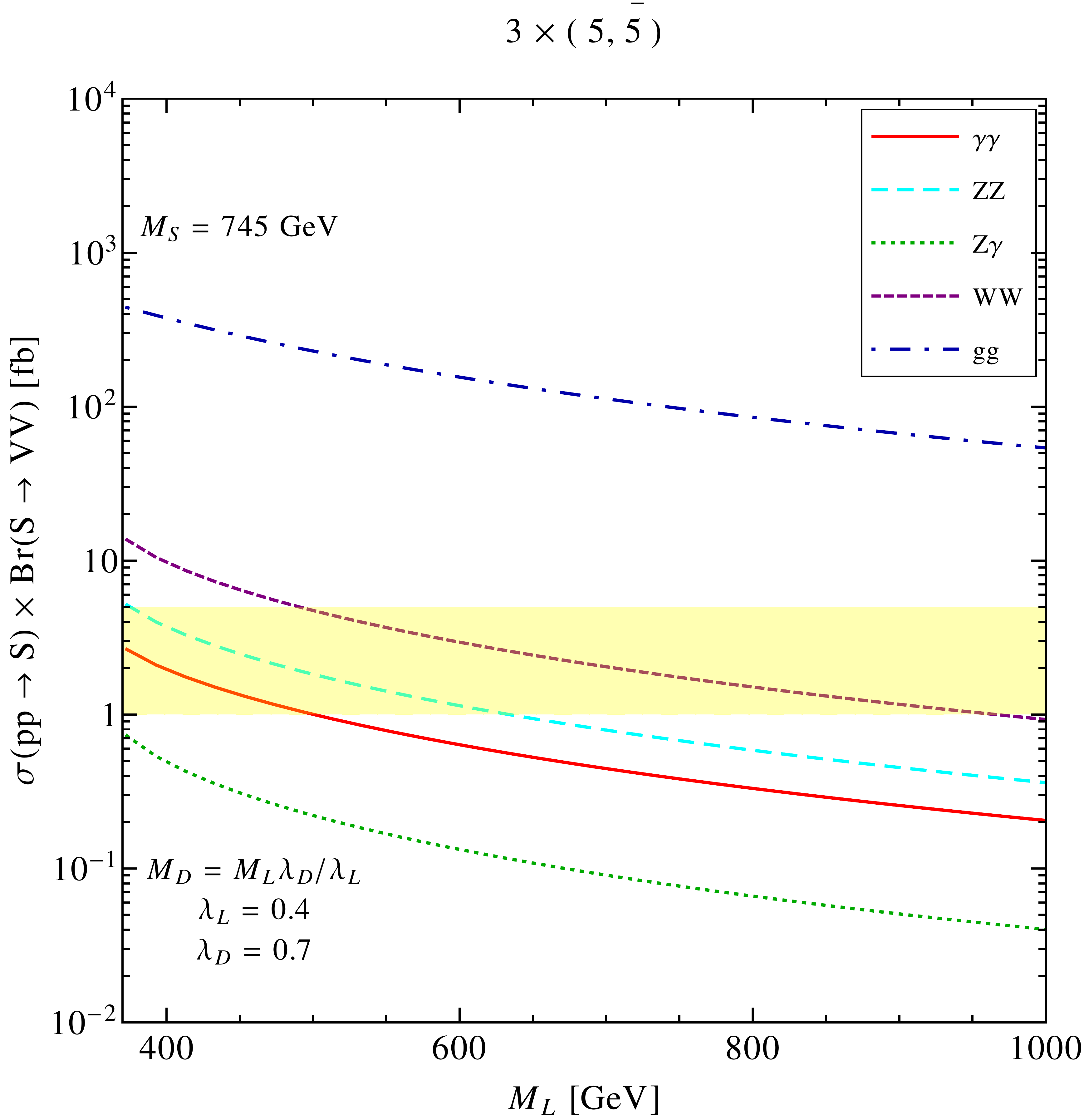}
\vspace{-0.3cm} \caption{[Left panel] The contours of $M_L$ and $M_D$ that fit
$\sigma_{\gamma \gamma} = 1-5$ fb for $M_S=745$ GeV. The black dashed line correspond to points belonging to the $SU(6)$ model under discussion.[Right panel] The corresponding cross-sections of $S$ decaying to various diboson channels as a function of $M_L$ with $M_D =M_L \lambda_D/\lambda_L$ as required by our $SU(6)$ model. The yellow shaded region shows the allowed values of $\sigma_{\gamma \gamma}$ for a narrow width resonance. $\lambda_L$ and $\lambda_D$ are set to 0.4 and 0.7 respectively for both plots. }
\label{fig:fitdata}
\end{figure}

In the left panel of Fig.~\ref{fig:fitdata} we present $\sigma_{\gamma \gamma}$ contours that fit the data,
for different values $M_L$ and $M_D$ belonging to generic $3 \times (5, \overline{5})$ models.
This figure clearly shows that the data can be fit for a range of values of $M_L$ and $M_D$. The reader should note that the
points belonging to the $SU(6)$ model under discussion are a subset of the generic $3 \times (5, \overline{5})$
points shown in the left panel. In our $SU(6)$ model $M_L/M_D = \lambda_L/\lambda_D$ is enforced 
since both $L$ and $D$ masses are generated by the $U(1)_X$ breaking VEV.
The points belonging to the $SU(6)$ model are shown by the black dashed line.
In the right panel of Fig.~\ref{fig:fitdata}, we show the cross-section times branching ratio of $S$
into various diboson channels as a function of the mass of $L$, keeping $M_D$ fixed at $M_L\lambda_D/\lambda_L$ as required by our $SU(6)$ model.
Evidently, the excess can be fit for $M_L \lesssim 500$ GeV. For this range of $M_L$ values,
the cross-sections in associated diboson channels are within current experimental limits.
We shall discuss the strongest of those limits in the subsequent paragraphs. 

\begin{table}[t]
\begin{center}
\setlength\tabcolsep{6pt}
\begin{tabular}{|c|c|c|c|c|c|c|c|c|}
\hline\hline
 & & $(M_L,M_D)$ & $\Gamma_S$ & $\sigma_{\gamma \gamma}$ &  $\sigma_{Z Z}$ &  $\sigma_{Z \gamma}$ & $\sigma_{WW}$ & $\sigma_{gg}$   \\
 & & [GeV] & [GeV] & [fb] & [fb] & [fb] & [fb] & [fb]  \\
 \hline
\multirow{2}{*}{SM + $3\times(5,\overline{5})$} & BP-1 & (374,655) & 0.03 & 2.61 & 5.06 & 0.71 & 13.4 & 437  \\
\cline{2-9}
\cline{2-9}
  & BP-2 & (500,875) & 0.02 & 1.00 & 1.82 & 0.22 & 4.73 & 229 \\
 \hline
 MSSM + $3\times(5,\overline{5})$ & BP-3 & (374,655) & 0.06 & 6.10 & 12.7 & 2.04 & 34.1 & 697 \\
\hline
\end{tabular}
\caption{Total decay width of $S$ and cross-sections in associated diboson final states for $M_S = 745$ GeV for
different BPs belonging to $3 \times (5, \overline{5})$. The choice of values of parameters $M_L, \, M_D$
are also shown, while fixing $\lambda_L = 0.4$ and $\lambda_D = 0.7$.
For the MSSM BP we choose $A_{f} = M_{\tilde{f}} = M_f$ ($f= D,L$) for simplicity. We also included $\alpha_{s}^4$ correction to $\Gamma_{gg}$, which increases it by a factor of $\sim 1.7$.
}
\label{tab:benchmarks}
\end{center}
\end{table}
In Table~\ref{tab:benchmarks}, we report the values of cross-sections in different channels along with the
total decay width for two BPs belonging to $3\times(5,\overline{5})$, setting $M_L/M_D = \lambda_L/\lambda_D$
as required by our $SU(6)$ model.
BP-1 and BP-2 respectively correspond to the best-fit and $2 \sigma$ lower limit of
$\sigma_{\gamma \gamma}$ needed for a narrow width resonance. Since the best-fit $\sigma_{\gamma \gamma}$ is achieved for $M_L \approx M_S/2$, we can not fit the $2 \sigma$ upper limit of it without introducing tree-level decay of $S$ into a pair of $L$. We note that using $M_L$ values of 374 GeV and 500 GeV for BP-1 and BP-2 respectively in the table above, the ratio of the corresponding diphoton production cross sections is 2.61 which is different from just $(500/374)^2=1.79$ due to the $M_L$ dependence of the loop functions. In addition, since $M_D$ values of BP-1 and BP-2 are not too high, their effect in this ratio can not be neglected either. The results presented in Table~\ref{tab:benchmarks} using decay width expressions of Eq.~(\ref{decay_widths}) agree with the numbers obtained from analytical expressions given in Refs.~\cite{diphoton:Strumia,diphoton:Zurek}.
The adoption of three copies of $(5, \overline{5})$ vector-like matter is well-motivated in a GUT context such as
$E_6\supset SU(6)$,  where the generations of new particles are in one-to-one correspondence with the SM generations.
Clearly, from Table~\ref{tab:benchmarks} and Fig.~\ref{fig:fitdata}, $3\times(5,\overline{5})$ fits the experimental data.
Further, we note that if the present $SU(6)$ model is supersymmetrized, then a large enhancement in $\sigma_{\gamma \gamma}$ is
possible due to loop contributions from superpartners of vector-like leptons and quarks.
This fact has been previously pointed out by Refs.~\cite{Dutta:SU5,Nilles}. In Table~\ref{tab:benchmarks}, we additionally
present a third benchmark (BP-3) that takes into account possible loop contribution from sleptons and squarks.
For simplicity, we assume $A_{f} = M_{\tilde{f}} = M_f$ for this supersymmetrized BP.
The inclusion of sparticles improves $\sigma_{\gamma \gamma}$ by a factor of $\sim 2.34$.


We now discuss constraints from a few associated diboson ($S \rightarrow W^+ W^-, \,  ZZ, \, Z \gamma$) final states
that arise from the decay widths presented in Eq.~(\ref{decay_widths}).
The $W^+W^-, \, ZZ, \, Z\gamma$ signals are estimated to occur with a rate comparable to that of the $\gamma\gamma$ channel,
being that they originate from the same set of couplings. Among these three weak-boson channels,
the $Z\gamma$ channel is the most stringent, and ATLAS~\cite{ATLAS:monophoton} constrains a monophoton signal to be less than $30$ fb at 13 TeV.
The two $3 \times (5,\overline{5})$ cases considered here clearly satisfy these bounds.

Next, we focus on the $gg$ channel in some detail, since it takes up a sizeable partial width in comparison to $\gamma\gamma$. CMS places the strongest $\sim 1.3$ pb bound on a 750 GeV $gg$ resonance at 13 TeV~\cite{ATLAS:dijet}. Evidently, our BPs survive the dijet bounds arising from the 13 TeV ATLAS analysis. 


Finally, in the case of a supersymmetrized model, we can additionally resolve the narrow-width problem by a possible
decay of $S$ into a pair of lightest supersymmetric particles (LSPs) with large width.
However, an invisible width sufficiently large to bring our Table~\ref{tab:benchmarks} BPs into compatibility with this
interpretation will be in slight tension with the monojet bounds~\cite{Dutta:SU5,Falkowski}. Such a conflict can easily
be avoided by promoting a candidate invisible final state into the `semi-invisible' regime, e.g. by decaying $S$
into a pair of next-to-LSPs (NLSPs), and thereafter allowing the NLSP to decay into the LSP and a relatively soft lepton.
This scenario may be realized efficiently via off-shell $Z^\ast/\tilde{l}^*$ decays associated with a kinematically narrow ($10-20$ GeV)
mass gap between the NLSP and LSP. Alternatively, the soft leptons can also be due to a slepton in between
the NLSP and LSP.  We refer the reader to Ref.~\cite{Dutta:SU5} for additional details.

\section{Conclusion}Vector like quarks and leptons with masses around the TeV scale are potentially beneficial
for explaining the resonant diphoton excess observed by CMS and ATLAS. However, this explanation triggers many additional questions,
such as whether the new scale is associated with any new symmetry, whether the new vector-like fermions and the SM fermions
belong to anomaly-free representations of any GUT group, and whether the scalar particle responsible for the 750 GeV resonance
can be economically associated with the new scale and the new symmetry multiplets housing the vector-like fermions.

In this paper, we made attempts to answer all of these questions in the context of an $SU(6)$ GUT model.
The vector-like fermions, along with the SM fermions, appear in the smallest anomaly free $15+\bar{6}+\bar{6}'$
representations of $SU(6)$, where $SU(6)$ breaks down to the SM$\times U(1)_X$ at the GUT scale.
Masses for the vector-like fermions are generated at the TeV scale where $U(1)_X$ is broken by a VEV of the
SM singlet field arising from ${\bar 6}_H$ and $21$. This singlet in the ${\bar 6}_H$ field is also responsible in turn for the observed resonance.
The dark matter arises from the SM singlet fermion residing in $\bar{6}$s, and is of Majorana type.
The SM fermions acquire masses at the electroweak scale.

We additionally demonstrated that a suitable gauge coupling unification is possible in this model,
and discussed the Yukawa couplings associated with the vector like fields in this model.
We also discussed the origin of neutrino masses.  The diphoton final states arise due to the 3 generations of
down type vector-like quarks and lepton doublets, where $M_L/M_D$ is fixed by a fixed point of the RGEs.
We used both 8 and 13 TeV diphoton excess results from ATLAS
and from CMS to calculate the masses and couplings associated with the new fields.  In addition to the diphoton
final states, we also expect $WW$, $WZ$, $ZZ$, and dijet final states.
We have carefully studied the final states associated with this model's
$SU(6)$ context, arriving at unique predictions that can be used to distinguish it at the LHC.

The 750 GeV diphoton excess has similarly been studied in the context of $U(1)'$ models by several groups,
e.g.~\cite{King:2016wep, Das:2015enc, Duerr:2016eme, Jiang:2015oms,Ko:2016lai,Staub:2016dxq}. 
but identification of the unifying $SU(6)$ GUT and the associated particle
content render this effort and its predictions different from other works.
In the present case, we have considered a non-supersymmetric $U(1)_X$ model
with fermionic vector-like particles.
In contrast with the supersymmetric $U(1)_X$ models, this construction does
not suffer from dimension-five proton decay via exchanges of scalar color triplets.
Also, vector-like masses are forbidden here by the $U(1)_X$ gauge symmetry,
and are generated dynamically only after $U(1)_X$ gauge symmetry breaking.
In Ref.~\cite{King:2016wep}, the supersymmetric $U(1)_N$ model has been studied. To avoid the
dimension-five proton decay problem,
the authors imposed a symmetry such as $Z_2^{qq}$ or $Z_2^{lq}$, which could
become subtle if one generation forms a complete fundamental representation of $E_6$.
Also, the doublets from vector-like particles are interpreted there as inert.
In Ref.~\cite{Das:2015enc}, vector-like particle masses are not forbidden by the $U(1)_X$ gauge
symmetry, and the model does not have an $E_6$/$SU(6)$ embedding. Similarly, in Ref.\cite{Duerr:2016eme},
an additional $U(1)_B$ has been considered without any unifying GUT symmetry.  In Ref.~\cite{Jiang:2015oms},
a supersymmetric $U(1)'$ model has been considered, which again cannot be embedded into $E_6$.
Ref.~\cite{Ko:2016lai} deals with a leptophobic $U(1)_X$ in the context of $E_6$. Finally, Ref.~\cite{Staub:2016dxq} analyzed and developed phenomenological tools by comparing all the $U(1)$ extension models proposed in the context of diphoton excess.

\begin{acknowledgments}
We thank Teruki Kamon and Zurab Tavartkiladze for helpful discussions. This work is supported in part by DOE grant numbers
DE-FG02-13ER42020 (B.D., T.G.),
the Mitchell Institute for Fundamental Physics and Astronomy (Y.G.),
Bartol Research Institute (I.G.), the  Rustaveli National Science Foundation  No. 03/79 (I.G.),
Natural Science Foundation of China grant numbers 11135003, 11275246, and 11475238 (T.L),
and National Science Foundation grant number PHY-1521105 (J.W.W.).

\end{acknowledgments}

\appendix


\section{\noindent{$SU(3)_C\times SU(2)_L \times U(1)_Y \times U(1)_X$ Model of $SU(6)$ from $E_6$ Embedding}}

$E_6$ has a subgroup $SU(6)\times SU(2)$, and the fundamental representation of $E_6$ decomposes
as
\be
{\bf 27} \longrightarrow ({\mathbf {\bar 6}}, {\mathbf 2})\oplus ({\bf 15}, {\mathbf 1})~.~\,
\ee
Thus, our $SU(3)_C\times SU(2)_L \times U(1)_Y \times U(1)_X$ model from $SU(6)$ can be embedded into $E_6$.
We consider
$
SU(6) \longrightarrow SU(5) \times U(1)_X~.~\,
$
And the generator of $U(1)_X$ is,
 $ T_{U(1)_X} = \frac{1}{2{\sqrt {15}}}  {\rm diag} \left(1,~1,~1,~1,~1,~-5\right) ~.$~\,
Thus, we obtain
\begin{eqnarray}
  {\bf 15} &\longrightarrow& ({\mathbf {\overline{10}}}, {\mathbf 2})\oplus ({\mathbf {5}}, {\bf -4})~,~\, \\
 {\mathbf {\bar 6}} &\longrightarrow& ({\mathbf {\bar 5}}, {\bf -1})\oplus ({\mathbf {1}}, {\mathbf 5})~,~\,
\end{eqnarray}
where the above $U(1)_X$ quantum numbers are $2{\sqrt {15}} Q_X$. In other words, the correct $U(1)_X$ charges
are the above $U(1)_X$ charges divided by $2{\sqrt {15}}$.

The $E_6$ gauge group can be broken as follows~\cite{Group,Hewett:1988xc}.
$E_6 \to\ SO(10) \times \ U(1)_{\psi} \to\ SU(5) \times\ U(1)_{\chi} \times\
U(1)_{\psi}~.$
The $U(1)_{\psi}$ and $U(1)_{\chi}$ charges for the $E_6$
fundamental ${\bf 27}$ representation are given in Table
\ref{E6charge}. The $U(1)'$ is one linear combination of
the $U(1)_{\chi}$ and $U(1)_{\psi}$,
$Q^{\prime} = \cos\theta \ Q_{\chi} + \sin\theta \ Q_{\psi}~.$
 For simplicity, we assume that the
other $U(1)$ gauge symmetry from the orthogonal linear combination
of the $U(1)_{\chi}$ and $U(1)_{\psi}$ is absent or broken at a
high scale. For the fundamental representation ${\mathbf 27}$ decomposition,
see  Table~\ref{E6charge}.

To realize the $U(1)_X$ gauge symmetry of $SU(6)$, we require that two singlets in Table~\ref{E6charge}
have the same $U(1)'$ charges. Thus, we obtain,
  $\cos\theta= -{\sqrt \frac{3}{8}}~.$
We present the $U(1)'$ charges in Table~\ref{E6charge}, and the $U(1)'$ charges are
indeed the same as our $U(1)_X$ charges.
Such kinds of $U(1)'$ models have been studied before~\cite{Langacker:2008yv, general, PLJW,
  Erler:2002pr, Kang:2004pp, Kang:2004ix, Kang:2009rd}. In particular, the $U(1)'$ gauge symmetry
in Ref.~\cite{PLJW} is the same as our $U(1)_X$ gauge symmetry from $SU(6)$, up to the overall
sign difference for the charges. However, in Ref.~\cite{PLJW}, the authors did not embed the $U(1)'$ model
into an $E_6$ model explicitly, and the gauge symmetry breaking
$E_6\rightarrow SU(3)_C \times SU(2)_L \times U(1)_Y \times U(1)'$ may be non-trivial.

\begin{table}[h]
\caption{Decomposition of the $E_6$ fundamental  ${\bf 27}$
representation under $SO(10)$ and $SU(5)$, and the $U(1)_{\chi}$,
$U(1)_{\psi}$ and $U(1)'$ charges.}
\begin{center}
\begin{tabular}{|c| c| c| c| c|}
\hline $SO(10)$ & $SU(5)$ & $2 \sqrt{10} Q_{\chi}$ & $2 \sqrt{6}
Q_{\psi}$ & $2 \sqrt{15} Q'$ \\
\hline
16   &   ${\mathbf{10} }$ & --1 & 1  & $2$ \\
            &   ${\mathbf {\bar 5}}$  & 3  & 1  & $-1$        \\
            &   ${\mathbf 1} $             & --5 & 1  & $5$         \\
\hline
       10   &   ${\mathbf 5}$    & 2  & --2 & $-4$         \\
            &   ${\mathbf {\bar 5}}$ & --2 &--2 & $-1$ \\
\hline
       1    &   ${\mathbf 1}$                  &  0 & 4 & 5 \\
\hline
\end{tabular}
\end{center}
\label{E6charge}
\end{table}

\end{document}